\documentclass[prl,aps,showpacs,
twocolumn,
nofootinbib,
floatfix,superscriptaddress]{revtex4}
\usepackage{graphicx}
\usepackage{epsfig}

\newcommand{\beq}{\begin{equation}}
\newcommand{\eeq}{\end{equation}}
\newcommand{\beqs}{\begin{eqnarray}}
\newcommand{\eeqs}{\end{eqnarray}}

\newcommand{\Tr}{{\rm Tr}}

\def\hbar{\hspace{0pt}\raisebox{1pt}{$-$} \hspace{-7pt} h}

\def\di{\mbox{d}}

\begin{document}
\title{Walking in the third millennium.}

\author{Maurizio Piai \thanks{email:piai@u.washington.edu }}
\affiliation{Department of Physics, University of Washington,
Seattle, WA 98195}

\date{September 11, 2006}


\begin{abstract}

Based on recent progress in the study of strong dynamics
in conformal field theories, I construct a simple, one family model
of dynamical electro-weak symmetry breaking. Non-perturbative effects 
can be computed systematically. The model  
 does not suffer from the well known, parametrically big,
phenomenological difficulties of traditional technicolor models,
is compatible with all present experimental data and
is testable at the LHC.

\end{abstract}


\maketitle

\section{Introduction}

Models of dynamical electro-weak symmetry breaking suffer
from well-known parametrically big phenomenological problems.
A possible solution
consists of assuming that the dynamics be very different from
QCD, with an energy range of quasi-conformal behavior at strong coupling
near the confinement scale and  the presence of 
large anomalous dimensions ({\it walking}~\cite{walking}). 
This provides a parametric enhancement of fermion masses,
and might soften
the problems with precision electro-weak data~\cite{precision}. 
But the lack of a reliable computational tool
for the non-perturbative 
effects has been a strong limitation on quantitative studies~\cite{AS}.

Some peculiar four-dimensional strongly interacting field theories,
based on  $SU(N_T)$ gauge theories at large-$N_T$,
admit a dual description in terms of a weakly coupled gravity theory in 
a higher dimensional space-time. This is the conjectured AdS-CFT correspondence~\cite{AdSCFT}.
It is the ideal tool for the description
of walking technicolor and allows to address the long-standing problem of
estimating the magnitude of non-perturbative effects on 
precision electro-weak parameters~\cite{MP}.

In this letter, I propose a very simple model of walking technicolor, 
formulated in the language of the AdS-CFT correspondence, in
which non-perturbative effects are calculable. 
The model is not complete. 
It lacks a dynamical explanation
for the hierarchical pattern of fermion masses. 
It also does not provide a 
UV-complete four-dimensional Lagrangian valid at energies above the
scale at which conformal symmetry is lost.

This simple one-family model  illustrates how a modern approach
to the physics of strong-interacting, conformal field theories,
leads very naturally to the solution of long-standing problems
of model-building,
and allows to construct realistic, calculable and testable models of walking technicolor.
Fine-tuning problems with the $\hat{S}$ and $\hat{T}$ parameters and the top
mass $m_t$  are easily softened by
one order of magnitude. Higher-order operators are
suppressed by the UV cut-off at which conformal symmetry is lost,
estimated to be in the $5-10$ TeV range.
The bounds from precision electro-weak physics can be satisfied even 
in the large-$N_T$ regime, in contrast to what suggested by perturbative estimates.
The low-energy phenomenology  is the one of the standard model (SM) with a heavy,
broad, strongly interacting and model-dependent scalar Higgs.
The model is testable:  a set of four spin-1
excited states have degenerate masses in the $2-4$ TeV range, with
strong couplings to the SM currents, and hence 
unsuppressed production and decay rates at the LHC.

\section{The Model.}

The energy window  above the electro-weak scale
is  described by a slice of $AdS_5$, 
 a five-dimensional space-time containing a  gravity 
background with the metric:
\beqs
\di s^2 &=& \left(\frac{L}{z}\right)^{2}\left( \eta_{\mu\nu}\di x^{\mu}\di x^{\nu}\,-\,\di z^2\right)\,,
\eeqs
where $x^{\mu}$ are four-dimensional coordinates, $\eta_{\mu\nu}$ the Minkoski 
metric with signature $(+,-,-,-)$, and $z$ is the extra (warped) dimension.
$L$ is the $AdS_5$ curvature.
Conformal symmetry is broken by  the boundaries 
$
L_0\,<\,z\,<\,L_1\,,
$
with $L_0>L$. $L_0$ and $L_1$ correspond to the UV and IR cut-offs of the
conformal theory. At energies above $1/L_0$ 
 the four-dimensional strong-interacting sector is no 
longer conformal, and the gauge coupling of the dual $SU(N_T)$ theory
runs toward
asymptotic  freedom. The 
SM with a  heavy, 
strongly interacting,
Higgs boson is recovered below $1/L_1$.

The field content in the bulk of the five-dimensional  model 
consists of a  complex scalar $\Phi$ transforming 
as a $(2,1/2)$ of the gauged  $SU(2)_L\times U(1)_Y$. 
The generators of $SU(2)_L$ are $T_i=\tau_i/2$ with $\tau_i$ the 
Pauli matrices.
The bulk action for $\Phi$ and the gauge bosons $W = W_i T_i$ 
and $B$  is
\beqs
{\cal S}_{5} &=& \int\di^4 x \int_{L_0}^{L_1}\di z\,\sqrt{G}\left[\frac{}{}
\left(G^{MN}(D_M\Phi)^{\dagger} D_N\Phi -M^2|\Phi|^2\right)\,\nonumber\right.\\
&&\left. \left(-\frac{1}{2}\Tr\left(W_{MN}W_{RS}\right)-\frac{1}{4}B_{MN}B_{RS}\right)G^{MR}G^{NS}\right]\,,
\eeqs
with the boundary terms given by
\beqs
{\cal S}_{4} &=& \int\di^4 x \int_{L_0}^{L_1}\di z \,\sqrt{G}\left[ \frac{}{}\delta(z-L_0)G^{\mu\rho}G^{\nu\sigma}\right.\\
&&\left[-\frac{1}{2}D\Tr\left[W_{\mu\nu}W_{\rho\sigma}\right]-\frac{1}{4}D
B_{\mu\nu}B_{\rho\sigma}\right]
\nonumber\\
&& -\delta(z-L_i)\,2\lambda_i \left(|\Phi|^2-\frac{\mbox{v}_i^2}{2}\right)^2\nonumber
\eeqs
where the $D_M\Phi$ is the covariant derivative and $i=0,1$.
In the action, $M^2=-4/L^2$ is a bulk mass term for the scalar, 
and $g$ and $g^{\prime}$ are
the (dimensionful) gauge couplings in five-dimensions.

Electro-weak symmetry breaking is induced by 
the VEV of $\Phi$:
\beqs
\langle \Phi \rangle &=& \frac{\mbox{v}(z)}{\sqrt{2}}\left(\begin{array}{c}
0\cr1\end{array}\right)\,.
\eeqs
 In the $\lambda_i\rightarrow +\infty$ limit,
the bulk equations admit the solution
\beqs
\mbox{v}(z)&=&\frac{\mbox{v}_1}{L_1^2}z^2\,=\,\frac{\mbox{v}_0}{L_0^2}z^2\,.
\eeqs
With this choice 
symmetry-breaking is triggered by a chiral condensate of dimension $d=2$,
i.e.  a bilinear fermionic condensate with anomalous dimension $\gamma=1$.

In the following, I focus on the spin-1 sector of the model in unitary gauge. 
The complete Lagrangian, including the appropriate gauge-fixing
terms, is discussed in~\cite{MP}.

\section{Electro-weak Phenomenology.}

I define:
\beqs
V^{M}&\equiv & \frac{g^{\prime}W_3^{M}+gB^{M}}{\sqrt{g^2+g^{\prime\,2}}}\,,\\
A^{M}&\equiv & \frac{g W_3^{M}-g^{\prime}B^{M}}{\sqrt{g^2+g^{\prime\,2}}}\,,
\eeqs
so that  the massless 
mode of $V^{\mu}$ is the photon, and the lightest
mode of $A^{\mu}$ is the $Z$ boson.

After Fourier transformation in the four-dimensional Minkoski coordinates:
\beqs
A^{\mu}(q,z)&\equiv&A^{\mu}(q)v_Z(z,q)\,,
\eeqs
and analogous for $W_{1,2}$ and $V$, 
where $q=\sqrt{q^2}$ is the four-dimensional momentum.
The bulk equations are:
\beqs
\partial_z\frac{L}{z}\partial_z v_i-\mu^4_{i} L z  v_i&=&-q^{2}\frac{L}{z}v_i\,,
\eeqs
where $i=v,Z,W$, with $\mu_v=0$,  $\mu^4_W=1/4g^2\mbox{v}_0^2/L^2$ and 
$\mu^4_Z=1/4(g^2+g^{\prime 2})\mbox{v}_0^2/L^2$.

The bulk equations can be solved exactly. The matrix of the polarizations $\pi_i(q^2)$
of the SM gauge bosons can be written in terms of the 
action evaluated at the UV boundary:

\begin{widetext}
\beqs
\frac{\pi_{+}}{{\cal N}^2}&=&Dq^2+\frac{\partial_zv_{W}}{v_{W}}(q^2,L_0)\,,
\\
\frac{\pi_{BB}}{{\cal N}^2}&=&
Dq^2 +\frac{g^2}{g^2+g^{\prime\,2}}\frac{\partial_zv_{v}}{v_v}(q^2,L_0)
+\frac{g^{\prime\,2}}{g^2+g^{\prime\,2}}\frac{\partial_zv_{Z}}{v_Z}(q^2,L_0)\,,
\\
\frac{\pi_{WB}}{{\cal N}^2}&=&
\frac{g g^{\prime}}{g^2+g^{\prime\,2}}\left(\frac{\partial_zv_{v}}{v_v}(q^2,L_0)
-\frac{\partial_zv_{Z}}{v_Z}(q^2,L_0)\right)\,,\\
\frac{\pi_{WW}}{{\cal N}^2}&=&
Dq^2 +
\frac{g^{\prime\,2}}{g^2+g^{\prime\,2}}\frac{\partial_zv_{v}}{v_v}(q^2,L_0)
+\frac{g^{2}}{g^2+g^{\prime\,2}}\frac{\partial_zv_{Z}}{v_Z}(q^2,L_0)\,,
\eeqs
\end{widetext}
where ${\cal N}$ is 
chosen to produce canonical kinetic terms in the
limit in which the heavy resonances  decouple.
The precision electro-weak parameters are defined as
\beqs
\hat{S}&\equiv&\frac{g_4}{g_4^{\prime}}\,\pi_{WB}^{\prime}(0)\,,\\
\hat{T}&\equiv&\frac{1}{M_W^2}\left(\pi_{WW}(0)-\pi_{+}(0)\right)\,,
\eeqs
where  $g_4^{(\prime)}$
 are the (dimensionless) 
gauge couplings of the SM in four-dimensions,
and where $\pi^{\prime}\equiv \di \pi/\di q^2$.

Taking the limit $L_0\rightarrow L$, and expanding  for small  $L_0\rightarrow 0$,
from~\cite{MP}:
\begin{widetext}
\beqs
\frac{\partial_zv_{v}}{v_v}(q^2,L_0)&=&q^2L_0\left(\frac{\pi}{2}
\frac{Y_0(qL_1)}{J_0(qL_1)}-\left(\gamma_E+\ln \frac{qL_0}{2} \right)\right)\,,\\
\frac{\partial_zv_{Z}}{v_Z}(q^2,L_0)&=&
L_0\left\{
\mu_Z^2\,-\,q^2\left[
\gamma_E+\ln(\mu_Z L_0)+\frac{1}{2}\psi\left(-\frac{q^2}{4\mu_Z^2}\right)-\frac{c_2}{2c_1}\Gamma\left(-\frac{q^2}{4\mu_Z^2}\right)
\right]\right\}\,,
\eeqs
where, after imposing Neumann boundary conditions in the IR, 
\beqs
c_1&=&2L\left(-1+\frac{q^2}{4\mu_Z^2},\mu_Z^2L_1^2\right)+L\left(\frac{q^2}{4\mu_Z^2},-1,\mu_Z^2L_1^2\right)\,,\\
c_2&=&-U\left(-\frac{q^2}{4\mu_Z^2},0,\mu_Z^2L_1^2\right)+\frac{q^2}{2\mu_Z^2}
U\left(1-\frac{q^2}{4\mu_Z^2},1,\mu_Z^2L_1^2\right)\,,
\eeqs
\end{widetext}
and where $\partial_zv_{W}/v_{W}=
\partial_zv_{Z}/v_{Z}(\mu_Z\rightarrow \mu_W)$. 

The  localized counterterm 
\beqs
D&=&L_0\left(\ln\frac{L_0}{L_1}+\frac{1}{\varepsilon^2}\right)\,
\eeqs
cancels the logarithmic divergences, and for ${\cal N}^2=\varepsilon^2/L_0$ 
all the dependence on $L_0$  disappears (at leading order in $L_0$), 
the limit $L_0\rightarrow 0$ can be taken,
and the model is renormalized, with finite  SM couplings
$g_4^{(\prime)\,2}= \varepsilon^{2}g^{(\prime)\,2}/L$.

Expanding for $\mu^2_ZL_1^2\ll1$, 
\beqs
\hat{S}&=&\varepsilon^2
\frac{1}{2e}\mu_W^4L_1^4\,,\\
\hat{T}&=&\frac{\varepsilon^2}{M_W^2}\left(\mu_W^2\tanh\frac{\mu_W^2L_1^2}{2}-
\frac{\mu_W^4}{\mu_Z^2}\tanh\frac{\mu_Z^2L_1^2}{2}\right)\,\\
&\simeq&\frac{\varepsilon^2}{M_W^2}\frac{\mu_W^4L_1^6}{24}(\mu_Z^4-\mu_W^4)\,,
\eeqs
with $e\simeq 2.7$.

The spin-1 spectrum consists of 
the massless photon, the $W$ and $Z$ gauge bosons,
and towers of their excited states. The four towers are essentially degenerate:
the splitting between the first excited state of  $W$ and $Z$ is parametrically smaller
than their mass, which for all practical purposes coincides with 
the mass $M_{\rho^0}=k/L_1$ of the first excited state of the photon, 
the lightest techni-$\rho$ resonance, where $k\in[2.4,4.7]$ grows with $\varepsilon$.
The mass of the  $W$ gauge boson is approximately given by
\beqs
M_W^2&\simeq&\varepsilon^2\left(\mu_W^2\tanh\frac{\mu_W^2L_1^2}{2}\right)\,\simeq\,\frac{1}{2}\varepsilon^2\mu_W^4L_1^2\,,
\eeqs
while $M_Z^2\simeq (g^2+g^{\prime\,2})/g^2M_W^2$.
Equivalently,
\beqs
M_W^2&=&\frac{1}{8}\varepsilon^2g^2\mbox{v}_1^2\left(\frac{L_0}{L_1}\right)^2
\,=\,\frac{1}{4}g_4^2\eta^2\,,
\eeqs
where $\eta$ is related to the Fermi decay constant $G_F$ by
\beqs
\eta^2&=&L\frac{\mbox{v}_1^2}{2}\left(\frac{L_0}{L_1}\right)^2\,=\,\frac{1}{\sqrt{2}G_F}\,\simeq\,(246 \,{\rm GeV})^2\,.
\eeqs

Substituting  in
the precision parameters yields:
\beqs
\hat{S}&\simeq&\frac{1}{e}M_W^2L_1^2\,=\,\frac{k^2}{e}\frac{M_W^2}{M_{\rho^0}^2},\\
\hat{T}&=&\frac{M_Z^2-M_W^2}{6\varepsilon^{2}}L_1^2
\,=\,\frac{k^2}{6\varepsilon^{2}}\frac{M_Z^2-M_W^2}{M_{\rho^0}^{2}}\,.
\eeqs

I take as indicative of the experimentally allowed ranges
(at the $3\sigma$ level):
\beqs
\hat{S}_{exp}&=& (-0.9\pm 3.9) \times 10^{-3}\,,\\
\hat{T}_{exp}&=& (2.0\pm3.0) \times 10^{-3}\,,
\eeqs
from~\cite{precision}. These bounds  are extrapolated to
the case of a Higgs boson with mass of $800$ GeV. The  comparison has to be 
done with some caution. The one-loop level SM analysis used
in the extraction of the bounds is not reliable for a heavy, strongly coupled Higgs,
the mass of which is not controllable in this model.

All SM fields are fundamental fields with no TC interactions,
hence localized on the UV-boundary.
With usual assignments to the representations of the SM gauge group, 
the mass of the top  originates after symmetry breaking
from the  (dimensionful) Yukawa 
coupling 
$-\, \delta(z-L_0)
y_u \bar{q_L} \tilde{\Phi} u_R$
where $\tilde{\Phi}=i\tau_2 \Phi$ and where for simplicity $L=L_0$.
The renormalized top-Yukawa in four dimensions 
$y_4\equiv y_uL_0/L_1\sqrt{L}$
reproduces the experimental value  $m_t\sim 175$ GeV
for $y_4\sim1/\sqrt{2}$.
At finite $L_0\sim L$, this requires:  
\beqs
\frac{y_u}{\sqrt{L}}&=&\frac{L_1}{\sqrt{2}L_0}\,.
\eeqs

\section{Naturalness and Experimental bounds.}

As long as $g^2/L$ is small the (tree-level) computations in the five-dimensional 
 gravity background should provide a good estimate of
 the non-perturbative effects in the dual four-dimensional conformal gauge theory.
 This is the large-$N_T$ limit.
 Since $g_4^2=\varepsilon^2g^2/L$ is the SM $SU(2)_L$ gauge coupling, 
 parametrically small 
 values for $\varepsilon$ imply parametrically big uncertainties. 
 Hence, I assume $\varepsilon>1/2$ ($g/\sqrt{L}<1.3$). 
 Numerically, $k(\varepsilon=1/2)\simeq 2.8$.
 For these values of $\varepsilon$,
the bounds on $\hat{T}$ are automatically satisfied if the bounds
 on $\hat{S}$ are. 
 
 The indicative bound on $\hat{S}$ implies:
 \beqs
 \frac{1}{L_1}& > & \frac{ M_W}{\sqrt{e \hat{S}_{\rm max}}}\,
 =\, 890\, {\rm GeV}\,.
 \eeqs
 The techni-$\rho$ mesons have mass $M_{\rho^0}\simeq2.5$ TeV.
 There are four of them, for all practical purposes degenerate in mass.
 The next excited states have masses bigger than $5$ TeV.
 With  small values of $g/\sqrt{L}$,
  the couplings of the lightest techni-$\rho$'s 
 to the SM currents are not suppressed,
 and hence they should be observable at the LHC.
 
 In order to gauge the amount of fine-tuning that might be 
 hidden in the definition of the
 renormalized couplings, it is convenient to look at the physical quantities
 at finite $L_0$ and $L$.
The bound on $\hat{S}$ is independent of $g$ and $L_0$.
From the expression for $\eta$ it translates into:
\beqs
\frac{\mbox{v}_1^2LL_0^2}{2}&=&\eta^2L_1^2\,<\,\left(\frac{1}{3.6}\right)^2\,.
\eeqs

The natural expectation for $\mbox{v}_1$ is
\beqs
\label{natural}
\mbox{v}_1&\simeq&\frac{2.4}{gL_1}\,,
\eeqs
leading  to the bound:
\beqs
\label{bound}
\frac{L}{g^2}\frac{L_0^2}{L_1^2}&<&\left(\frac{1}{6}\right)^2\,.
\eeqs

Before commenting on this bound, Eq.~(\ref{natural}) 
deserves to be explained. $L_1$ is the confinement scale,
at which the symmetry-breaking condensate of the $SU(N_T)$ theory 
$\mbox{v}_1$ forms.
$\mbox{v}_1$ has dimensions $3/2$, which on dimensional grounds explains 
the extra factor of $g$, since these are the only two parameters relevant 
on the IR-brane.
The numerical coefficient is deduced from the real-world QCD relation 
$\sqrt{2}g_{\rho}f_{\pi}=M_{\rho}$, with $g_{\rho}=g/\sqrt{L}$ and $M_{\rho}=2.4/L_1$. 
The value of $g_{\rho}\simeq  6$ is
extracted from the 
first coefficient of the large-$q^2$ expansion of the vector-vector 
correlator:
$L/g^2=N_c/12\pi^2$.
Eq.~(\ref{natural}) should not be taken too literally.
It only gives a size of the {\it natural} expectation for the value
of the symmetry-breaking condensate evaluated at the confinement scale.

Eq.~(\ref{bound}) is the main result of this paper. 
At large-$N_T$ (small $g$), the bound on $\hat{S}$ is
satisfied thanks to the parametric scale separation between $L_0$ and $L_1$,
which provides a suppression mechanism to compensate for the smallness of $g^2/L$.
For $g^2/L\sim 1$, this implies $1/L_0 \sim 6/L_1 \sim 5.3$ TeV, 
high enough to provide a sufficient suppression of higher order operators,
and  enough to suppress FCNC transitions in a complete ETC model~\cite{APS}.
The value of the top Yukawa is in this case somewhat large $y_u/\sqrt{L} \sim 4$,
but still natural.
From a phenomenological perspective,  walking  
makes the axial and vectorial contributions to  $\hat{S}$ cancel each other,
so that the bounds are satisfied even if the spin-1 resonances 
have unsuppressed couplings to the SM currents.

This has to be contrasted with what happens in  a small-$N_T$ 
QCD-like technicolor theory, in which there is no substantial cancellation
between axial and vectorial contribution to $\hat{S}$, and the bounds
are satisfied because
 a large value of $g^2/L$ parametrically suppresses
the techni-$\rho$ decay constants.
From the perspective of the searches at the LHC, the scenario
described here is very appealing. A more
quantitative study is needed, but going to large-$N_T$
renders the production and decay rates bigger
than in a QCD-like model,
improving on the feasibility of the search itself.

The perturbative estimate of $\hat{S}$ completely misses
the  $(L_0/L_1)^2$ suppression factor coming from walking.
The limit case $g/\sqrt{L}\sim 1.3$ corresponds roughly
to a $SU(N_T)$ theory with critical  number $N_d\sim 2N_T$~\cite{ATW} of
techni-fermions transforming on the fundamental of both $SU(2)_L$
and $SU(N_T)$ for $N_T\sim 8$. In this case, the perturbative estimate
is:
\beqs
\hat{S}_p&=&\frac{\alpha}{4\sin^2\theta_W}\frac{N_dN_T}{6\pi}\,
\sim\,0.06\,,
\eeqs
compared with $\hat{S}\simeq 0.003$, obtained for  $1/L_0\sim 5$ TeV.

\vspace{1.0cm}
\begin{acknowledgments}
I am grateful to T.~Appelquist
for encouraging this study
and for useful discussions.
This work is partially supported by the DOE grant
DE-FG02-96ER40956.
\end{acknowledgments}


\end{document}